\documentclass[useAMS,usenatbib]{mn2e}
\usepackage{graphicx}
\usepackage{psfig}
\usepackage{epsf}
\title[Short GRBs in old populations]
{Short GRBs in old populations: Magnetars from WD-WD mergers}

\author[A.J.~Levan et al.]{Andrew J.~Levan$^{1,2}$\thanks{email: levan@star.herts.ac.uk},
 Graham A. Wynn$^{3}$, Robert Chapman$^{1}$,
Melvyn B. Davies$^{2}$, \and Andrew R. King$^{3}$,
Robert S. Priddey$^{1}$ and Nial R.Tanvir$^{1}$\\
$^{1}$Centre for Astrophysics Research, University of
Hertfordshire, College Lane, Hatfield AL10 9AB, UK.\\
$^{2}$Lund Observatory, Box 43, SE--221 00, Lund, Sweden.\\
$^{3}$Department of Physics and Astronomy, University of Leicester,
Leicester, LE1~7RH, UK. \\
}

\begin{document}

\date{Accepted 2006 January 13. Received 2006 January 12; in original 
form 2005 December 17}

\pagerange{\pageref{firstpage}--\pageref{lastpage}} \pubyear{2002}

\maketitle

\label{firstpage}

\begin{abstract}
Recent progress on the nature of short duration $\gamma$-ray bursts
has shown that a fraction of them originate in the local universe. These
systems may well be the result of giant flares from soft gamma-repeaters
(highly magnetized neutron stars commonly known as magnetars).
However, if these neutron stars are formed
via the core collapse of massive stars then it would be expected 
that the bursts should originate from predominantly young stellar populations, 
while correlating the positions of BATSE short bursts with structure in the
local universe reveals a correlation with all galaxy types, including
those with little or no ongoing star formation. 
This is a natural outcome if, in addition to magnetars formed
via the core collapse of massive stars they also form via
Accretion Induced Collapse following the merger of two white dwarfs, 
one of which is magnetic. We investigate this
possibility and find that the rate of magnetar production via WD-WD mergers
in the Milky Way is comparable to the rate of production 
via core-collapse. However, while the rate of production of magnetars by core
collapse is proportional to the star formation rate, the rate
of production via WD-WD mergers (which have long lifetimes) is proportional
to the stellar mass density, which is concentrated in 
early-type systems. Therefore magnetars produced via WD-WD mergers may produce 
SGR giant flares which can be identified with early type galaxies. We also
comment on the possibility that this mechanism could produce 
a fraction of the observed short duration burst population at higher redshift.
\end{abstract}


\section{Introduction}

Soft $\gamma$-repeaters (SGRs) are thought to be highly magnetized 
neutron stars with rotational periods of seconds and strong 
($>10^{14}G$) magnetic fields (see e.g. Kouveliotou et al. 1998).
They undergo frequent outbursts 
and occasional giant flares (Hurley et al. 1999;  Palmer et al. 2005).
These giant flares are sufficiently
bright to be observed in external galaxies out to large distances 
($>50$ Mpc), and would be observable as a short duration
$\gamma$-ray burst (Hurley et al. 2005). However, the paucity of observed giant flares in 
our own galaxy has so far precluded observationally based determinations of either
their luminosity function or rate.

Observations of recently localised short duration GRBs 
have shown them to be associated with a variety of host galaxy types at typical
redshifts of $z \sim 0.2$ (e.g. Gehrels et al. 2005; Fox et al. 2005; Berger et al. 2005).
Their energetics, and location in a variety of host
galaxies, including those with little or no ongoing star formation point
to an origin in the merger of two compact objects, either NS-NS,
or NS-BH (Bloom et al. 2005; Hjorth et al. 2005). 

However, if at least some fraction of short duration GRBs are due to 
SGRs in external galaxies then it may be expected that some correlation
between the positions of bursts and the locations of galaxies in the local
universe
would be seen. Tanvir et al. (2005 - hereafter T05) have recently reported such 
a correlation, indicating that up to $\sim 25$\% of short duration
GRBs originate in the local universe (within 100 Mpc). Intriguingly this analysis
shows a correlation with all galaxy types, and is strongest when 
restricted to relatively early-type galaxies (Sbc and earlier). SGRs are thought
to be formed in the core collapse of massive stars and due to relatively short
lifetimes ($\sim 10^4$ years; e.g. Kouveliotou 1999) would naturally be located in predominantly
star forming galaxies, while essentially none should be seen in
ellipticals. Two possibilities present themselves: the first
is that rather than representing SGRs the correlations reported by T05 
are due to a different progenitor class, such as low-luminosity
neutron star mergers. The second possibility is that SGRs are not 
created exclusively by core collapse events and can be found in older
stellar populations. This second option is the subject of this paper. 

Here we consider an alternative model for the creation of SGRs, and thus 
potentially GRBs: namely SGRs which are created via the Accretion
Induced Collapse (AIC) of white dwarfs to neutron stars (e.g.  Nomoto \& Kondo 1991).
Such an AIC may
occur in WD-WD mergers (King, Pringle \& Wickramasinghe 2001),
or rarely  in binaries with more massive
main sequence companions although the precise outcome
of WD-WD mergers remains uncertain, and may produce either
a SN Ia or an AIC.
Once the magnetar has been created its evolution will proceed in an analogous
manner to those created via core collapse. 

\section{Making Magnetars from white dwarf mergers}
King, Pringle \& Wickramasinghe (2001) first suggested that the merger of two white
dwarfs (one or more of which was highly magnetic) may
result in the production of a magnetar via
AIC. Observational 
evidence for such a channel can be found from the highly magnetic 
white dwarf REJ 0317-853, which
appears to be the result of the merger of two white dwarfs with
a total mass marginally below the Chandrasekhar limit. For
more massive systems a magnetar may be formed. 

However, the required field strengths for the white dwarfs are
very large. Upon collapse the new magnetic field is
given by $B_{NS} = B_{WD} (R_{WD}/R_{NS})^2$. For typical 
white dwarf and neutron star parameters this implies that
white dwarf B-fields of several hundred MG are necessary for magnetar  creation. 
Such fields are relatively rare, but do exist within the magnetic
white dwarf population. Figure 1 shows the distribution of
magnetic fields in isolated white dwarfs and in magnetic 
CVs (Wickramasinghe \& Ferrario 2000; Schmidt et al. 2003; Norton et al. 2004;
Vanlandingham et al. 2005). As can be seen a small fraction ($\sim 10$\%) achieve the
required field strengths. Thus it is plausible that
the accretion induced collapse of a white dwarf during the merging
process could create a magnetar.

\begin{figure}
\begin{center}
\resizebox{8truecm}{!}{\includegraphics{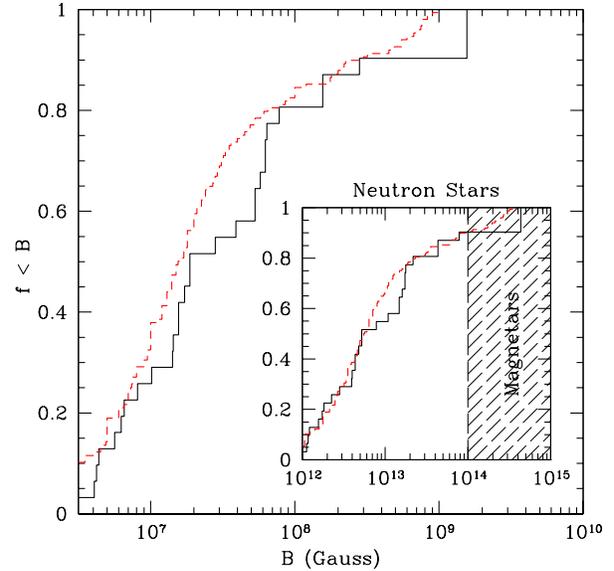}}
\end{center}
\caption{The distribution of white dwarf magnetic fields seen in mCVs
(black line - 33 stars taken from Norton et al. 2004) and isolated white dwarfs (red dashed line
- 148 stars with $B> 2$MG, taken from Wickramasinghe \& Ferrario 2000, Schmidt et al. 2003;
Vanlandingham et al. 2005).
For magnetic CVs we have converted from magnetic moment to B-field
assuming that R=$6 \times 10^8$cm 
(the radius of a mean mass (0.95 M$_{\odot}$) magnetic white dwarf). A
more detailed description including the effects of the white dwarf mass
radius relation is given in Section 3.
The inset shows
the fields following collapse into a neutron star of radius $1 \times 10^6$ cm. 
Magnetars are defined as having $B>10^{14}$G.}
\label{f1}
\end{figure}

\begin{figure}
\begin{center}
\resizebox{8truecm}{!}{\includegraphics{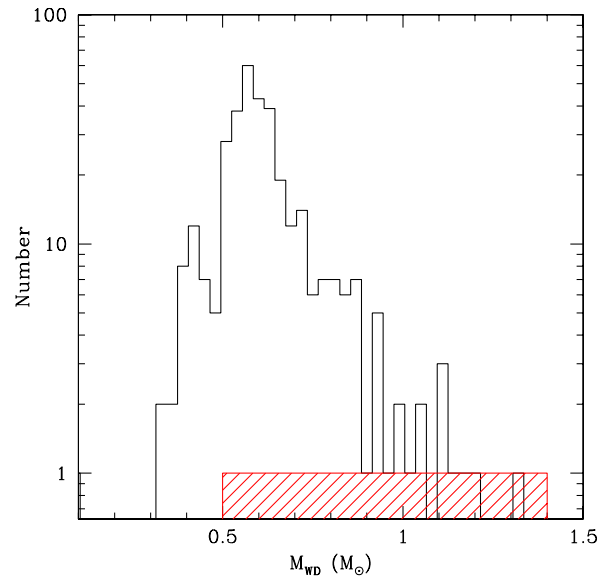}}
\end{center}
\caption{The mass distributions assumed for this work. The
non-magnetic white dwarf distribution has been taken from Liebert,
Bergeron \& Holberg (2005) while the magnetic distribution is assumed to be flat
over the mass range of $0.5 < M < 1.4$, as is shown
in the hatched area. The relative numbers of 
white dwarfs in each population are those assumed here. }
\label{f1}
\end{figure}

\begin{figure}
\begin{center}
\resizebox{8truecm}{!}{\includegraphics{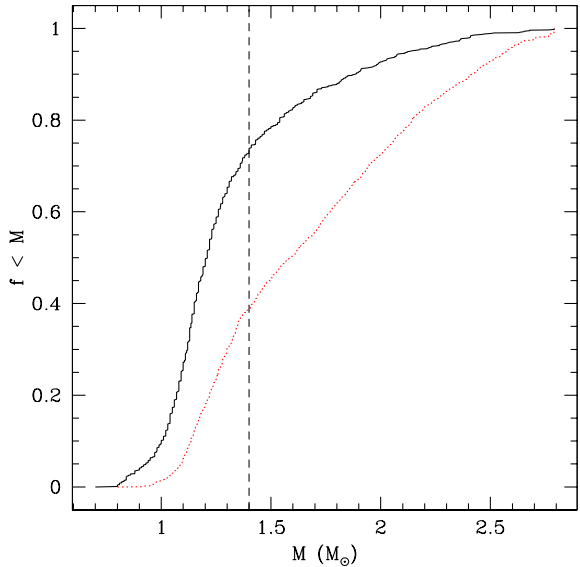}}
\end{center}
\caption{The distribution of the combined masses of mergers 
of CO white dwarfs, where
each component is picked to favour mass ratios close
to unity as described in section 3. The red dotted line
shows the mass distribution of double degenerate systems which contain
a magnetic white dwarf. As can be seen it is charaterised by a higher mean
mass, due to the differing shape of the magnetic white dwarf mass function. 
The dashed vertical line represents the Chandrasekhar mass.}
\label{f1}
\end{figure}

\section{The rate of formation of magnetars from WD-WD mergers}

We now proceed to estimate roughly the rate of formation of SGRs via 
the WD-WD channel within the Milky Way. Initially we construct a mass distribution containing
both magnetic and non-magnetic carbon-oxygen white dwarfs (Figure 2). The mass
function for the non-magnetic systems is taken from Liebert, Bergeron \& Holberg (2005),
exhibiting a broad structure with a peak at $\sim 0.57$ M$_{\odot}$. 
The mass function for highly magnetic white dwarfs is markedly different from this, and
for this work is considered to be flat  between
0.5 and 1.4 M$_{\odot}$ with a mean of 0.95 $M_{\odot}$ (this
is consistent with current observations of magnetic white dwarfs - 
Liebert, Bergeron \& Holberg 2003; Nalezyty
\& Madej 2004). The fraction
of magnetic white dwarfs (with B $>2MG$) within this model is 9\%,
again consistent with recent observations (e.g. Liebert, Bergeron \& Holberg 2003).
{For each white dwarf we obtain a radius using the formulation of Nauenberg (1972),
and, for magnetic white dwarfs calculate the B-field that would be
formed upon collapse to a neutron stars of radius $1 \times 10^{6}$cm.
We only consider carbon-oxygen (CO) white dwarfs since although
He-He systems are the dominant double degenerate population
by number they are of less interest as sources of magnetars, 
since they will rarely exceed the Chandrasekhar mass.
Detailed population synthesis calculations by Nelemans et al. (2001)
find that $\sim$ 25\% of the WD-WD systems formed consist of
two CO white dwarfs.

We do not consider the detailed evolutionary pathways that lead to 
the production of the WD-WD binaries but do pick white dwarfs
from the mass distribution such that the observed mass ratios are
consistent with observations (Maxted et al. 2002). To this
end we pick the first white dwarf at random from the entire
mass distribution but then pick the secondary from a gaussian
function, centered on the mass of the first WD with a
FWHM of 20\% of this mass. This preferentially produces
binaries with mass ratios close to unity, in agreement
with observations of double degenerate systems.

We find that the fraction of double degenerate systems  which form
above the Chandrasekhar mass is $\sim$ 25\% of
our population (or $\sim 6$\% of the total WD-WD
population if CO-He and He-He systems are also considered). 
As we  do not attempt detailed population synthesis calculations here, we
instead normalize the double degenerates created with masses 
$>$ 1.4 M$_{\odot}$ against the rate obtained via population
synthesis (Nelemans et al. 2001), so that the rate of such mergers
is $3 \times 10^{-3}$ yr$^{-1}$.
Of  this population $\sim$ 40\% contain
at least one magnetic white dwarf. Although
this appears to be a larger fraction than might naturally be anticipated it is 
due to the fact that magnetic white dwarfs exhibit a higher mean mass, so
that the distribution of WD-WD binaries containing a magnetic white
dwarf is biased towards higher masses (see Figure 3 - at masses above $\sim 1$ M$_{\odot}$
the mass function of magnetic white dwarf contributes approximately
the same number of stars as the non-magnetic population). In 
approximately 10\% of the double degenerate systems
(with M$>$1.4M$_\odot$) the magnetic fields
are sufficiently strong to form a magnetar should an AIC occur
upon merger.

\begin{figure}
\begin{center}
\resizebox{8truecm}{!}{\includegraphics{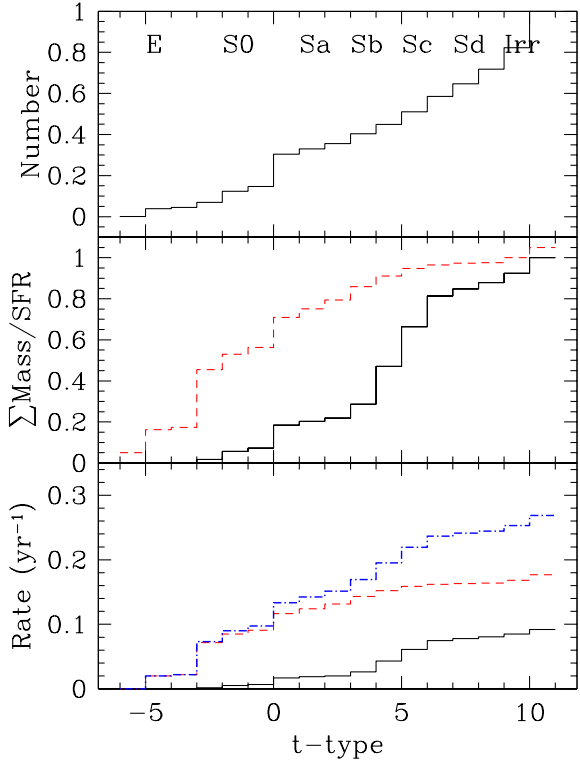}}
\end{center}
\caption{
Top: The distribution of different T-types within the Third Reference Catalogue of
Bright Galaxies
(RC3 - de Vaucouleurs et al. 1995) with
 $v<2000$ km s$^{-1}$. The T-type distribution runs along the Hubble sequence
between -6 (Elliptical) and 11 (Irregular), the Hubble type for a given 
T-type is indicated in the top pannel. 
 Middle: The cumulative distribution of stellar mass (red dashed) and star formation (black solid)
 within the same velocity cut, the model for stellar mass and star formation assumed 
 to create this is described in Appendix A. As can
be seen the much of the stellar mass is maintained in early-type systems, while
star formation takes place predominantly in later systems. The lower panel shows
the extrapolated rates of SGRs which follow (red dashed line) stellar mass
at a rate of $3.5 \times 10^{-4}$ yr$^{-1}$ $\times$ ($M /10^{11}$ M$_{\odot}$)
and (solid black line) follow the star formation rate as
$1 \times 10^{-4}$ yr$^{-1} \times$ (SFR /  M$_{\odot}$  yr$^{-1}$). Within the considerable uncertainties,
the rates of each channel within the local universe are comparable and thus
we may expect to see a correlation between the locations of short bursts
and all galaxy types.  The blue dot-dashed line shows the cumulative
rate of SGR formation via both channels. As can be seen the rate of formation
in earlier-type galaxies (T-type 4 and earlier), accounts for $\sim$ 70\% of the
total rate.}
\label{f1}
\end{figure}

\section{Discussion}
The currently popular picture for the production of magnetars 
is that they result from the core collapse of stars
more massive than 40 M$_{\odot}$ (e.g. Gaensler et al. 2005). The rate of production of 
such stars within the galaxy is $\sim 6 \times 10^{-4}$ yr$^{-1}$ (Podsiadlowski 
et al. 2004).
The rate of SGRs (created by both WD-WD and core collapse channels) 
can also be estimated based on the observed number within our galaxy. Given
a lifetime of $10^4$ years and the observed number within the MW (4) the estimated
rate is $\sim 4 \times 10^{-4}$ yr$^{-1}$.
The location of the observed galactic SGRs within the galactic plane and the association in
some cases with young stellar clusters makes it reasonable to assume that
all of the galactic SGRs have been created via core collapse, although
this is not certain. The total rate is also a lower limit since it is possible that some
galactic SGRs have been quiescent and have thus far evaded detection.

Nelemans et al. (2001) estimate a rate of mergers of white dwarf binaries
with masses higher than the Chandrasekhar mass of $3 \times 10^{-3}$ yr$^{-1}$. 
Thus if $\sim 10\%$ of these systems create a magnetar then we expect a galactic rate
of magnetar production via WD-WD mergers of $3 \times 10^{-4}$ yr$^{-1}$.
This this is comparable to the 
rate inferred from core collapse.

 The rate of core collapse events within the local universe should trace the
 massive star formation rate (SFR), while the rate of SGRs formed via WD-WD mergers
 will better trace the stellar mass density, since the merger of
 the white dwarf pair can occur several Gyr after the star formation which
 created the white dwarf progenitors.  In early-type systems the
 rate of production of SGRs via WD-WD mergers will be significantly larger
 than via core collapse. For example, in an early type galaxy,
the SFR may be $\sim$ 0.1 M$_{\odot}$ yr$^{-1}$, while the stellar
mass may be similar to the Milky Way. In such systems the rate of
production via core collapse would be $\sim 1 \times 10^{-5}$ yr$^{-1}$, while
the rate of production via WD-WD mergers would be unchanged at
$\sim 3 \times 10^{-4}$ yr$^{-1}$, a factor of ten higher than the 
core collapse rate.

Given the SFR and stellar mass of the MW  (following the 
calculations of Nelemans et al. (2001), we assume 
SFR = 4 M$_{\odot}$ yr$^{-1}$ and $M_{disk} = 8 \times 10^{10}$ M$_{\odot}$)
we estimate
the rate of production by WD-WD mergers and by core collapse
events to be $\Re_{WD-WD} \approx 3.5 \times 10^{-4}$ yr$^{-1}$ $\times$ ($M / 10^{11}$ M$_{\odot}$)
and $\Re_{CC} \approx 1 \times 10^{-4}$ yr$^{-1}$ $\times (SFR$ / (M$_{\odot}$ yr$^{-1}$)).
An extrapolation of this to the local universe is shown in Figure 4. The rate
 of SGR production via core collapse and
WD-WD mergers is comparable within the uncertainties. Thus,
SGRs produced via WD-WD mergers would naturally predict a
correlation between the locations of short GRBs and galaxies of
all types 
in the local universe. 
Indeed, the total rate of SGR production 
in earlier-type galaxies (T-type 4 and earlier, where T05 find
their strongest correlation) is predicted to be
$\sim 0.2$ yr$^{-1}$, or 70\% of the total SGR production, providing
a natural explanation of this result. 
Some enhancement of the rate of WD-WD mergers may also be seen in globular 
clusters where the systems can be formed dynamically as well as via
the normal binary evolution channels. Although this enhancement
is probably only modest (see e.g. Davies \& Benz 1995) it would 
primarily act to enhance the rate of mergers in 
very early-type systems, where a large number of globular clusters are
found.

The predicted total rate of $\sim 0.3$ yr$^{-1}$ combined with
the lifetime of SGRs of $1 \times 10^4$ yrs implies that within
a velocity cut of $<2000$km s$^{-1}$ there exist $\sim$ 3000 SGRs. 
The results of T05 indicate that $\sim$ 3 short GRBs
per year occur within this volume. If this is the case then it implies that
the rate of SGR giant flares is approximately one per SGR every
1000 years (i.e. within the MW where there are 4 SGRs a giant flare
of the magnitude of that seen from SGR 1806-20 in December 2004 would
be expected every 250 years). This is not unreasonable since
it is thought that giant flares similar to that seen in SGR 1806-20 might
release $\sim 10\%$ of the magnetar dipole energy, therefore 
this rate fits well with the estimated lifetime of SGRs of $10^4$ years.

It is interesting to note that a primary piece of evidence 
used to argue that the recently located short burst population
is the result of compact binary (neutron star or black hole)
mergers is the association of short GRBs with
all galaxy types, {\it including} those with little or no ongoing star formation. 
This immediately rules out any connection with events requiring
ongoing star formation. Under the standard
assumption that SGRs are produced via core collapse, this observation
can also be used to argue that SGRs cannot be responsible for
the recent well localised short duration bursts. However it should be noted that
under our proposed model we would expect to locate bursts in all 
galaxy types. Indeed the galaxy type fractions for WD-WD mergers may be 
similar to
those of SN Ia, or may even be biased to typically earlier type galaxies if
the SN Ia population is represented by two progenitor classes (one
of which is enhanced in younger populations e.g. Mannucci, Della Valle \& Panagia (2005)). 
Given the long lifetimes of the WD-WD dwarf systems we would
not expect to find them in, or near regions of massive star formation, and
in this sense their locations in host galaxies may be similar to
NS-NS mergers. However the dynamics of their formation will
differ from NS-NS and NS-BH binaries (in the sense that WD-WD
systems will not have natal kicks as neutron stars do) and so
the location of bursts around their
hosts may enable the distinction between different models. 

There are limits to the energetics which can feasibly 
be achieved via SGR giant flares. Those observed within 
the Milky Way have achieved luminosities up to $10^{46}$ ergs, although
Hurley et al (2005) comment that it may be possible to achieve 
luminosities a factor of 100 greater. Nonetheless the most luminous
short duration bursts with energies of 10$^{50}$ ergs would
appear to be beyond the energies which can be created via SGRs,
implying that two populations of bursts are still required. Ultimately
observations of SGRs in nearby galaxies may enable the
determination of their luminosity function and fix the rate
at which they occur. 

\section{Conclusions}
We have presented a mechanism for the production of magnetars via 
Accretion Induced Collapse following the merger of two white dwarfs
in a tight binary. Considering the observed mass and magnetic field 
distributions for white dwarfs it appears that the rate of formation
of SGRs via this process in the local universe is comparable to the production rate
via the core collapse channel (although possibly slightly lower
than the core collapse channel in the Milky Way). In early-type
galaxies the WD-WD mechanism is the dominant producer of SGRs,
while in late type galaxies most SGRs are produced via core collapse. 
Thus, it would be expected that SGR flares appearing as short duration
GRBs may be found from galaxies of all types and this naturally explains
the correlation seen between the locations of BATSE short duration
bursts and early-type galaxies in the local universe.

\section*{Acknowledgments}
We thank the referee for a rapid and constructive report, and for
comments on the white dwarf mass-radius relation. 
AJL \& NRT are grateful to PPARC for postdoctoral and senior research
fellowship awards. AJL also thanks the Swedish Institute for support while
visiting Lund. Astrophysics research at Leicester and Hertfordshire is funded
by a PPARC rolling grant. RC thanks the University of Hertfordshire
for a studentship. MBD is a Royal Swedish Academy Research Fellow
supported by a grant from the Knut and Alice Wallenberg Foundation.
ARK gratefully acknowledges a Royal 
Society--Wolfson Research Merit Award.

\appendix
\section{Mass and Star Formation in the Local Universe}
In order to estimate the rate of SGR production within different galaxy types
in the local universe, we have created a crude model
which estimates the star formation and mass of galaxies within the Third 
Reference Catalogue of Bright Galaxies, RC3 (de Vaucouleurs et al. 1995).

We initially cut the RC3 catalogue including only galaxies with
$v < 2000$ km/s, roughly 28 Mpc. We then cross correlate the 
positions of these galaxies with the 2MASS extended source
catalogue to obtain B-K colours for each of the galaxies. 
We are then able to estimate the mass of each individual galaxy
using the relation given by Mannucci et al. (2005)

\begin{equation}
\log{{\left( M/ L_K \over M_{\odot} / L_{\odot}\right)}} = 0.212(B-K) - 0.959
\end{equation}

Having ascertained the mass we now  derive rough estimates for
the star formation contained within each galaxy T-type.
We estimate this based on the
mean SFR of galaxies of a given T-type (from Shane \& James 2002).
The rates we assume are E (0 M$_{\odot}$ yr$^{-1}$), S0 (0.2 M$_{\odot}$ yr$^{-1}$),
Sa (0.25 M$_{\odot}$ yr$^{-1}$), Sb (0.5 M$_{\odot}$ yr$^{-1}$),
Sc (1.4 M$_{\odot}$ yr$^{-1}$), Scd (1.1 M$_{\odot}$ yr$^{-1}$),
Sd (0.7 M$_{\odot}$ yr$^{-1}$), Irr (0.15 M$_{\odot}$ yr$^{-1}$).
While these numbers only crudely estimate the star formation rate for a given
galaxy the total star formation rate obtained via this assumption is
in agreement with the extrapolation of the local SFR to $v<2000$ km s$^{-1}$
(Gallego et al. 1998).

\label{lastpage}

\end{document}